\documentstyle[twocolumn,prb,aps,epsf]{revtex}
\begin{document}
\def \beq{\begin{equation}}
\def \eeq{\end{equation}}
\def \beqarr{\begin{eqnarray}}
\def \eeqarr{\end{eqnarray}}

\twocolumn[\hsize\textwidth\columnwidth\hsize\csname @twocolumnfalse\endcsname

\draft

\title{
Spin mapping, phase diagram, and collective modes in double layer quantum 
Hall systems at $\nu=2$
}

\author{ 
Kun Yang
}

\address{
Condensed Matter Physics 114-36, California Institute of Technology,
Pasadena, California 91125\\
and Institute for Theoretical Physics, University of California, Santa
Barbara, California 93106}
\date{\today}
\maketitle

\begin{abstract}
An exact spin mapping is identified to simplify the recently proposed
hard-core boson description (Demler and Das Sarma, Phys. Rev. Lett., to be
published) 
of the bilayer
quantum Hall system at filling factor 2. 
The effective spin model describes an easy-plane ferromagnet subject to an
external Zeeman field.
The phase diagram of this effective model is determined exactly
and found to agree with the approximate calculation of Demler and Das Sarma, 
while the 
Goldstone-mode spectrum, order parameter
stiffness and Kosterlitz-Thouless temperature 
in the canted antiferromagnetic phase
are computed approximately.
\end{abstract}

]

Quantum Hall systems have attracted prolonged interest, because they
not only exhibit the extremely interesting quantized Hall effect, 
but also prove to be the ideal arenas to study many other important 
subjects, such as Anderson localization, 
Luttinger liquid behavior, itinerant magnetism, to name a few.
Recently, much attention
has been focusing on multicomponent quantum Hall systems.\cite{gm} Such 
components may be the spins of electrons which are not frozen out when the 
external magnetic field is not too strong, or layer indices in multi-layered
system. Novel physics such as even-denominator fractional quantum Hall
states,\cite{suen,eisenstein} spin-ferromagnetism and topological
excitations,\cite{sondhi} and spontaneous
inter-layer coherence\cite{wenzee,ezawa,murphy,yang}
have been found.

Very recently, a novel canted antiferromagnetic phase has been proposed 
theoretically in bilayer quantum Hall systems at filling factor $\nu=2$,
based primarily on Hartree-Fock calculations.\cite{zheng,brey}
There is some experimental
support for the existence of such a phase.\cite{pellegrini}
Later, an interesting attempt\cite{demler} was made to go beyond the
Hartree-Fock theory. In this attempt, one first truncate the Hilbert space
in such a way that the sum of the
number of electrons for a given Landau orbital is always two
for both layers, so
that the total charge degree of freedom is not allowed to
fluctuate. The problem is further simplified by keeping only
two lowest energy states for a given orbital. With these simplifications
and mapping the reduced Hamiltonian onto hard-core bosons on a lattice, 
Demler and Das Sarma\cite{demler} were able to obtain a phase diagram that
agrees well with Hartree-Fock in the absence of disorder, and 
predict some novel properties of the system with disorder.
The advantage of this formalism is that certain local correlations are treated
non-perturbatively, and effects of disorder are easy to incorporate. 
Thus it is complementary to Hartree-Fock.

The purpose of the present paper is to show that
the hard-core bosons introduced in Ref. [\onlinecite{demler}]
form an exact Schwinger boson representation of an effective spin-1/2 system. 
In terms of the spin-1/2 operators, the reduced Hamiltonian of 
Ref. [\onlinecite{demler}] takes a particularly simple form; it describes
an easy-plane ferromagnet subject to a
magnetic field along the $z$ direction, while
the $XY$ plane is the easy plane. The advantage of this spin-mapping
is that a lot of insight from, and theoretical techniques developed for
spin models (which are {\em not} available in the original
hard-core representation) can be applied to the present problem. In the
spin model, it is clear
that the phase diagram obtained in Ref. [\onlinecite{demler}] based on
approximate
variational wave functions that only satisfy the hard-core condition {\em 
in average}, is actually {\em exact}, {\em within the reduced Hamiltonian}. 
More importantly, the standard spin-wave theory 
applied to the present model
yields the exact collective mode spectrum for the singlet and triplet phases
(again within the reduced Hamiltonian), as well as the approximate 
Goldstone mode spectrum and order parameter stiffness in the canted phase.
From the order parameter stiffness we can also estimate the Kosterlitz-Thouless
temperature for the canted phase.

We begin with the reduced hard-core boson Hamiltonian, Eq. (4) of Ref.
[\onlinecite{demler}]:
\beqarr
H&=&E_t\sum_it_i^{\dagger}t_i+E_v\sum_iv_i^{\dagger}v_i
-{J\over2}\sum_{\langle ij\rangle}t_i^{\dagger}t_it_j^{\dagger}t_j\nonumber\\
&-&{J\over 4}(\cos\theta+\sin\theta)^2\sum_{\langle ij\rangle}
(t_i^{\dagger}v_iv_j^{\dagger}t_j+t_j^{\dagger}v_jv_i^{\dagger}t_i).
\label{reduced}
\eeqarr
Here $i$ is the index of a complete and orthogonal set of
localized intra-Landau level orbitals which form a
square lattice. $t_i^{\dagger}$ and $v_i^{\dagger}$ are boson creation 
operators which create {\em a pair} of electrons in orbital $i$;
the pair created by $t_i^{\dagger}$ have their spins
polarized along the external magnetic field 
to take advantage the Zeeman
energy, while the pair created by $v_i^{\dagger}$ occupy the symmetric 
combination of the two states in two layers to minimize the tunneling
energy, and form a spin singlet. These are two low-energy local states
being kept.
Since the number of electrons is fixed to be
2 for each orbital, they must satisfy the hard-core condition
\beq
t_i^{\dagger}t_i+v_i^{\dagger}v_i=1.
\label{hardcore}
\eeq
The parameters $E_t$, $E_v$, $J$ and $\theta$ are\cite{demler}
\beqarr
E_t&=&-H_z,\\
E_v&=&\epsilon_c/2-\sqrt{\Delta^2_{SAS}+\epsilon_c^2/4},\\
J&=&e^2/(16\sqrt{2\pi}\epsilon\ell),\\
\theta
&=&\arctan[\epsilon_c/(2\Delta_{SAS}+2\sqrt{\Delta^2_{SAS}+\epsilon_c^2/4})],
\eeqarr
where $H_z$ is the Zeeman splitting, $\epsilon$ is the dielectric constant of
the system, $\ell$ is the magnetic length, and $\epsilon_c=(e^2/\epsilon\ell)
\sqrt{\pi/2}(1-e^{-d^2/\ell^2}Erf[d/(\sqrt{2}\ell)])$ is the local charging
energy, in which
$d$ is the layer spacing and $Erf$ is the error function.

The constraint Eq. (\ref{hardcore}) implies that the $t$ and $v$ hard core
bosons form a Schwinger boson representation\cite{sakurai}
of a spin-1/2 operator, for 
each lattice site.
We may define the spin-up state to be the state in which a $t$ boson
is present, and the spin-down state to be the one with $v$ boson present.
The spin-1/2 operators are related to the hard-core boson operators in the
following way:
\beqarr
S_i^z&=&(t_i^\dagger t_i-v_i^\dagger v_i)/2,\\
S_i^+&=&t_i^\dagger v_i,\\
S_i^-&=&v_i^\dagger t_i.
\eeqarr

It is important to remember that these effective spin-1/2 degrees of freedom 
correspond to the singlet and spin-polarized electron pair states, and are not
directly related to the physical spins of the individual electrons.
In terms of these effective spin-1/2 operators, the reduced Hamiltonian
Eq. (\ref{reduced})
takes a particularly simple form:
\beq
H=-B_z\sum_iS_i^z-J_\parallel\sum_{\langle ij\rangle}(S_i^xS_j^x+S_i^yS_j^y)
-J_\bot\sum_{\langle ij\rangle}S_i^zS_j^z,
\label{ham}
\eeq
in which
\beqarr
B_z&=& JZ/4+E_v-E_t,\\
J_\parallel&=&J(\sin\theta+\cos\theta)^2/2,\\
J_\bot&=&J/2,
\eeqarr
where $Z$ is the coordination number of the lattice and for a square 
lattice, $Z=4$. 
Since $J_\parallel > J_\bot > 0$, Eq. (\ref{ham}) describes an
easy-plane ($XY$ plane) ferromagnet, in the presence of a Zeeman field along
the $z$ direction, $B_z$. Again, $B_z$ is an effective field that couples to
these effective spins, which should not be confused with the physical 
magnetic field. 

The existence of three phases of the original system is
particularly easy to understand in terms of the spin model (\ref{ham}). When
$B_z$ is large and positive, all the effective
spins are pointing up in the ground state
($|\Psi_\uparrow\rangle=|\uparrow\uparrow\cdots\uparrow\rangle$), 
which means 
all the pairs are in the $t$ state. This corresponds to the fully 
spin-polarized
ferromagnetic (FPF)   
phase. When $B_z$ is large and negative, all the effective
spins are pointing down 
($|\Psi_\downarrow\rangle=|\downarrow\downarrow\cdots\downarrow\rangle$), 
which means
all the pairs are in the $v$ state, this corresponds to the spin singlet (SS)
phase. In between, however, the effective spins would like to develop
a spontaneous polarization in the $XY$ plane, to take advantage of the
$J_\parallel$ coupling (since it is stronger than $J_\bot$). This state
with a spontaneously broken $XY$ or U(1) symmetry corresponds to the
canted antiferromagnetic (CAF) phase.
These three phases are schematically illustrated in Fig. 1.

The phase diagram of (\ref{ham}) may be determined {\em exactly}, because 
the spin polarized (in either up or down direction) states are exact eigen 
states of the Hamiltonian, 
due to the fact that $S^z_{tot}=\sum_iS_i^z$ is a conserved
quantity. So are the single magnon excitations on top
of them.
When $|\Psi_\uparrow\rangle$ is the ground state, the exact single
magnon spectrum is
\beq
E({\bf k})=B_z+2J_\bot-J_\parallel(\cos k_xa+\cos k_ya),
\eeq
where $a=\sqrt{2\pi}\ell$ is the lattice constant
for the square lattice.\cite{note} This FPF phase becomes
unstable against the CAF phase when the spectrum becomes soft at $k=0$,
namely $B_z+2J_\bot-2J_\parallel=0$, or
\beq
E_v-E_t + J(1-\sin 2\theta)=0.
\eeq
This determines the exact phase boundary separating the FPF and CAF phases,
{\em within the model} (\ref{ham}). The phase boundary separating
the CAF and SS phases may be obtained similarly:
\beq
E_v-E_t + J(\sin\theta + \cos\theta)^2=0.
\eeq
\begin{figure*}[h]
\centerline{\epsfxsize=9cm
\epsfbox{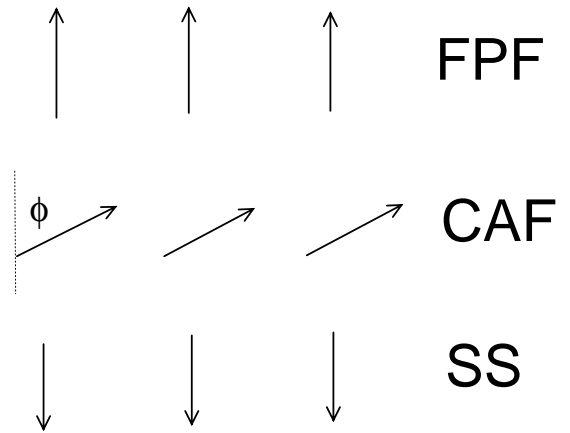}
}
\caption{
Configurations of effective spins in the three phases. All spins are pointing
up in the fully spin-polarized (FPF), and pointing down in the spin singlet
(SS) phase. In the canted antiferromagnetic (CAF) phase, the spin polarization
is tilted away from the $z$ direction by an angle $\phi$, and a component in
the $XY$ plane is spontaneously developed.
}
\label{fig1}
\end{figure*}
These agree 
with the results of Ref.
[\onlinecite{demler}], although in Ref.
[\onlinecite{demler}] it was not clear the phase boundaries obtained were 
exact within the reduced Hamiltonian, as variational wave functions 
that only satisfy the hard-core condition {\em on average} were used.

In the CAF phase, it is no longer possible to obtain the exact ground state
or magnon excitation spectrum, due to the presence of quantum fluctuations. 
In this case, we may perform an approximate spin-wave analysis. As a first
step, consider the following variational wave function, that describes the
state in which all the effective spins are still fully polarized, with 
the polarization tilted away from the $z$ direction by an angle $\phi$
(see Fig. 1):
\beqarr
|\phi\rangle&=&\prod_i[|\uparrow\rangle_i
\cos(\phi/2)+|\downarrow\rangle_i\sin(\phi/2)]\nonumber\\
&=&\prod_i[\cos(\phi/2)t^\dagger_i+\sin(\phi/2)v^\dagger_i]|vacuum\rangle.
\eeqarr
This variational wave function (with $\phi$ being the variational parameter)
is different from that used in Ref.\onlinecite{demler} and has the advantage
that the hard-core condition is satisfied exactly, in the original
hard-core boson representation.
The optimal $\phi$, which minimizes $\langle\phi|H|\phi\rangle$,
is found to satisfy 
\beq 
\sin\phi(B_z-2\Delta J\cos\phi)=0,
\label{angle}
\eeq
in which $\Delta J=J_\parallel-J_\bot=(J/2)\sin(2\theta)$.
The FPF and SS phases correspond to the trivial solutions $\phi=0$ and
$\phi=\pi$, while in the CAF phase we have 
\beq
\phi=\cos^{-1}[B_z/(2\Delta J)].
\label{phi}
\eeq
The phase boundaries can also be determined from the relative stability of
these solutions, which agree with the above spectrum analysis.

In order to perform a spin-wave analysis, we need to rotate the spin
quantization axis so that it is parallel to the spin 
polarization of $|\phi\rangle$,
after which the Hamiltonian reads
\beqarr
&H&=-B_z\sum_i(\cos\phi S_i^z+\sin\phi S_i^x)
-J_\parallel\sum_{\langle ij\rangle}{\bf S}_i\cdot{\bf S}_j
\nonumber\\
&+&\Delta J\sum_{\langle ij\rangle}(\cos\phi S_i^z+\sin\phi S_i^x)\cdot
(\cos\phi S_j^z+\sin\phi S_j^x).
\eeqarr
Now use the standard Holstein-Primakoff transformation\cite{manousakis}
to map spin operators to boson operators:
\beqarr
S_i^z&=&S-n_i, \nonumber\\
S_i^-&=&\sqrt{2S}f_S(n_i)a_i^\dagger,\nonumber\\
S_i^+&=&\sqrt{2S}f_S(n_i)a_i,
\eeqarr
where 
$a_i^\dagger$ is the boson creation operator for site $i$, $n_i=a_i^\dagger
a_i$, $S=1/2$ is the size of the effective spin, and 
\beq
f_S(n)=\sqrt{1-{n\over2S}}.
\eeq
Keep terms that are linear 
or quadratic in boson operators in $H$, we obtain the linearized 
spin wave Hamiltonian (up to a constant):
\beqarr
H_{LSW}&=&2J_\parallel\sum_ia_i^\dagger a_i+{1\over 4}\Delta J\sin^2\phi
\sum_{\langle ij\rangle}(a_i^\dagger a^\dagger_j+a_ja_i)
\nonumber\\
&-&({1\over 2}J_\parallel-{1\over 4}\Delta J\sin^2\phi)
\sum_{\langle ij\rangle}(a_i^\dagger a_j+a_j^\dagger a_i).
\eeqarr
The terms that are linear in $a$ cancel due to the condition Eq. (\ref{phi}).
This boson Hamiltonian may be diagonalized by first going to momentum space
and then perform a Bogliubov transformation. The diagonalized Hamiltonian
takes (up to a constant) the following form:
\beq
H_{LSW}=\sum_{\bf k}E_{\bf k}b_{\bf k}^\dagger b_{\bf k},
\eeq
where the Bogliubov boson $b$ is related to $a$ by
\beq
a_{\bf k}=\cosh\alpha_{\bf k}b_{\bf k}+\sinh\alpha_{\bf k}b_{-{\bf k}}^\dagger,
\eeq
in which
\beq
a_{\bf k}={1\over \sqrt{N}}\sum_ie^{-i{\bf k}\cdot {\bf r}_i}a_i,
\eeq
and the angle $\alpha_{\bf k}$ satisfies 
\beq
\tanh2\alpha_{\bf k}=-\Delta_{\bf k}/\epsilon_{\bf k},
\eeq
where 
\beqarr
\Delta_{\bf k}&=&[(\Delta J/2)\sin^2\phi](\cos k_xa+\cos k_ya),\nonumber\\
\epsilon_{\bf k}&=&2J_\parallel-
[J_\parallel-(\Delta J/2)\sin^2\phi](\cos k_xa+\cos k_ya).
\eeqarr
The spin wave spectrum is
\beq
E_{\bf k}=\sqrt{\epsilon_{\bf k}^2-\Delta_{\bf k}^2}.
\eeq
Taking the long wave length limit $k\rightarrow 0$, we find
a linear spectrum:
$E_{\bf k}=c|{\bf k}|$ expected for a Goldstone mode, with the
Goldstone mode (spin wave) velocity
\beq
c=a\sin\phi\sqrt{J_\parallel\Delta J}
={Ja\sin\phi\over 2}\sqrt{\sin 2\theta(1+\sin 2\theta)}.
\eeq

Quantum fluctuations of spin-waves reduces the magnetization in the CAF
phase. Within the linear spin-wave theory, the magnetization per unit cell
at $T=0$ is
\beq
M={1\over 2}-{1\over N}\sum_{\bf k}\langle a_{\bf k}^\dagger a_{\bf k}\rangle
=1-{1\over 2N}\sum_{\bf k}{1\over \sqrt{1-(\Delta_{\bf k}^2/
\epsilon_{\bf k}^2)}}.
\eeq
The susceptibility along $z$ direction is
\beq
\chi_z={\partial M_z\over \partial B_z}={\partial (M\cos\phi)\over
\partial B_z}=\cos\phi{\partial M\over \partial B_z}+{M\over 2\Delta J},
\eeq
in which we used Eq. (\ref{phi}).

The important quantity, order parameter stiffness $\rho_s$, is related to
$\chi_z$ and $c$ via\cite{halperin}
\beq
\rho_s=\chi_zc^2/a^2.
\eeq
It is straightforward to calculate $\rho_s$ numerically, and some numerical
data are presented in Fig. 2.

Assuming the short-distance physics at finite temperature to be similar
to that of the classical $XY$ model on a square lattice,\cite{zheng}
the Kosterlitz-Thouless transition temperature may be estimated to be
\beq
T_{KT}\approx 0.9 \rho_s.
\eeq
Thus in these systems, the typical $T_{KT}$ is of order $
T_{KT}\sim 0.2 J\sim 0.005 {e^2\over \epsilon\ell} \sim 0.5K$. This is 
qualitatively consistent, although somewhat smaller, than the Hartree-Fock
estimate.\cite{zheng} The fact that the present estimate gives a smaller
number may be due to fluctuations effects that have been taken into account
in the reduced Hamiltonian (which were left out in Hartree-Fock).

\begin{figure*}[h]
\centerline{\epsfxsize=9cm
\epsfbox{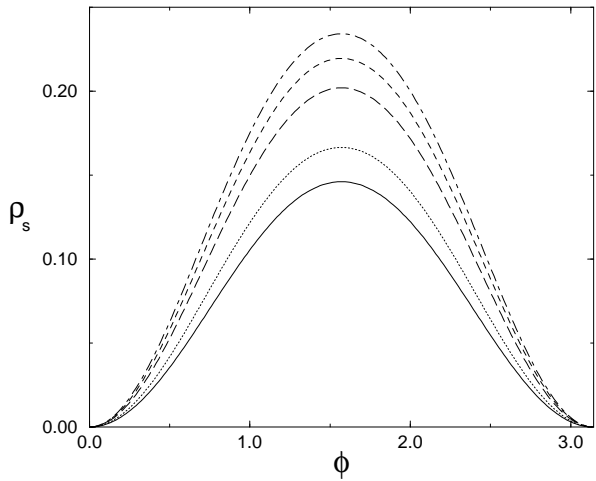}
}
\caption{
$\rho_s$ (in unit of $J$) as a function of $\phi$, for several different
$\theta$'s. From bottom up, $\theta=\pi/32, \pi/16, \pi/8, \pi/6, \pi/4$.
$\rho_s$ is invariant under $\theta\Rightarrow \pi/2-\theta$, or
$\phi\Rightarrow \pi-\phi$.
}
\label{fig2}
\end{figure*}

We emphasize that charge fluctuations have been left out in the reduced 
effective spin
Hamiltonian studied here, and it describes neutral excitations 
of the system {\em only}.
In particular, the skyrmions and
vortices of the effective spin model are charge-neutral.
This is very different from the bilayer systems at filling factor 1, where
all vortices must carry a half-integer charge.\cite{yang,ym} On the other
hand, the present system can support charged skyrmions and
vortices, which are being
studied.\cite{ez,paredes,dd}

The author has benefited greatly from stimulating discussions with Sankar
Das Sarma, Eugene Demler, and Luis Brey, while participating the ITP
programme on {\em Disorder and Interactions in Quantum Hall and Mesoscopic 
Systems}. This work was supported by the
Sherman Fairchild foundation (at Caltech), and NSF grant
PHY-9407194 (at ITP).


\begin{references}
\bibitem{note} The anisotropy in the spectrum is an artifact of the
lattice regularization of short-distance physics. However isotropy is 
restored in the long-wave length limit in all three phases, as expected.
\end{references}
\end{document}